\documentclass{sigchi}

\usepackage{balance}  
\usepackage{graphics} 
\usepackage{times}    
\usepackage{url}      

\makeatletter
\def\url@leostyle{%
  \@ifundefined{selectfont}{\def\UrlFont{\sf}}{\def\UrlFont{\small\bf\ttfamily}}}
\makeatother
\usepackage{cite}

\def\pprw{8.5in}
\def\pprh{11in}

\usepackage{multirow}

\usepackage{color}
\usepackage[]{graphicx}

\usepackage{listings}
\usepackage[pdftex]{hyperref}
\hypersetup{
pdftitle={M3I: Framework for Context-Based Mobile Multimodal Interaction},
pdfauthor={@@@},
pdfkeywords={multimodality; multimodal interaction; context awareness; framework; Android;},
bookmarksnumbered,
pdfstartview={FitH},
colorlinks,
citecolor=black,
filecolor=black,
linkcolor=black,
urlcolor=black,
breaklinks=true,
}

\begin{document}

\title{Supporting Mobile Multimodal Interaction with a Rule-Based Framework}

\numberofauthors{1}
\author{
  \alignauthor Andreas M\"{o}ller$~^{1}$, Stefan Diewald$~^{1}$, Luis Roalter$~^{1}$, Matthias Kranz$~^{2}$\\
    \affaddr{$^{1}$~Technische Universit\"{a}t M\"{u}nchen, Arcisstra{\ss}e 21, 80333 Munich, Germany}\\ 
    \affaddr{$^{2}$~Universit\"{a}t Passau, Innstra{\ss}e 43, 94032 Passau, Germany}\\
    \email{andreas.moeller@tum.de, stefan.diewald@tum.de, roalter@tum.de, matthias.kranz@uni-passau.de}
}


\maketitle

\begin{abstract}
Multimodality can make (especially mobile) device interaction more efficient.
Sensors and communication capabilities of modern smartphones and tablets lay the technical basis for its implementation.
Still, mobile platforms do not make multimodal interaction support trivial. Building multimodal applications requires various APIs with different paradigms, high-level interpretation of contextual data, and a method for fusing individual inputs and outputs.
To reduce this effort, we created a framework that simplifies and accelerates the creation of multimodal applications for prototyping and research.
It provides an abstraction of information representations in different modalities, unifies access to implicit and explicit information, and wires together the logic behind context-sensitive modality switches.
In the paper, we present the structure and features of our framework, and validate it by four implemented demonstrations of different complexity.
\end{abstract}




\section{Introduction and Motivation}
An emerged concept in the context of (mobile) interaction is \emph{multimodality} \cite{oviatt99}, i.e., using more than one modality (simultaneously or sequentially) for input and output.
This covers not only tactile, voice or gesture-based interaction, but also more implicit information, which is gathered from, e.g., the multitude of sensors that are nowadays available in mobile devices.
Particularly in mobile settings, individual unimodal interaction modes can suffer from limitations (e.g., the \emph{small screen} problem), so that multimodality may have a synergetic effect and may contribute to more efficiency \cite{oviatt99}.
Furthermore, human perception and communication uses all senses, so that multimodal interaction is considered more natural \cite{jokinen2006user} and thus more intuitive \cite{watanuki1994analysis}.

Multimodality support is beneficial in various individual uses cases. One example is in-car interaction \cite{DBLP:conf/mc/DiewaldMRK12,diewald2012mobilinet,diewald2013gameful}, where auditory in- and output and gestures \cite{pfleging2012} support hands-free interaction. One-handed interaction is e.g.~supported by distance-based input methods like DistScroll \cite{kranz2005distscroll}.

Often, the requirements for choosing a certain modality depend on the context, e.g., time, location, social setting, or security demands \cite{riedl2013momm}.
As an example for context-driven modality settings, Alice may not want to be notified on emails on weekdays between 10 PM and 6 AM; and Bob may mute his private phone at work, and always switch to hands-free mode in the car.
For each of those exemplified situations, a certain modality change (either on input or output side) occurs.
While these contextual modality switches mostly have to be performed manually (e.g., muting the phone before the meeting begins), 
they often occur in recurring situations, which are often based on rather simple rules. 
Most of the context types that play a role for modality decisions could even be inferred by the mobile device itself.

However, today's solutions address rather particular problems (take e.g.~the iPhone's \emph{Do Not Disturb} mode that pauses notifications in a definable period of time), which are either not sufficiently flexible, or have to be bought as additional software for each particular problem.
Making use of multimodality is still cumbersome for mobile software developers, although the necessary functionality is present in current devices. Yet, accessing and fusing contextual and sensor data in a meaningful way to create context-sensitive applications that feel ``smart'' is still a considerable effort.
End users do not have the possibility at all to create context-based behavior fitted to their personal needs.

Besides context-based switching of output channels, multimodality also allows new input methods and paradigms.
Novel interaction modalities have shown superior performance in e.g.~task completion time than traditional methods \cite{moeller2012mobimed} or enable novel scenarios in different domains such as indoor navigation \cite{moeller2012mum}, health and fitness \cite{moeller2012trainer}, or education and edutainment \cite{holleis2006playing,MobiDicsVideo2011}. However, the adoption of multimodality in the real world is still cautious, both from user and programmer side.

In this work, we present the \emph{M3I} (\emph{M}obile \emph{M}ulti\emph{M}odal \emph{I}nteraction) Framework \cite{muc2014m3i}, to our knowledge the first framework to support rich, context-driven multimodal interaction in mobile applications in a fast, holistic and simple manner. It allows researchers and application developers to integrate various modalities and to set up the decision logic for switching between them based on simple rules, all with a handful lines of code.
Explicit interaction is supported as well as implicit, context-driven behavior. No special setup and development tools are required, and the framework is fully extensible thanks to its modular structure.

The paper is organized as follows.
We conducted a focus group and expert interviews to deeper investigate current limitations and wishes, both from a user and a software developer point of view.
We briefly discriminate our framework from related research in multimodality and context awareness toolkits.
Building upon the requirements analysis, we then present the \emph{M3I} framework and its components, illustrated by code examples. We validate the claim that the \emph{M3I} framework drastically simplifies the support of multimodal interaction in mobile applications as well as the integration of context information. We show this at the example of several different use cases that demonstrate the broad applicability of the framework.
Finally, we discuss possible enhancements of the system for future work.

\section{Requirements Analysis}

As an initial step, we conducted a focus group and expert interviews to identify requirements both from a user's and a developer's perspective.

\subsection{User's Perspective: Focus Group}
The focus group was conducted with six participants (5 males, 1 female) between 24 and 30 years (average age: 27). Four participants were research assistants; two were students. Four of them owned an Android smartphone and two had an iPhone. All participants rated their smartphone expertise as “high” or “very high”. In a guided discussion, we investigated current usage and adoption of multimodality, and elicited unsatisfied needs and expectations. The focus group took about one hour and was audio-recorded. In the following, we summarize the most important results and design implications.
\subsubsection{Current usage of input and output modalities}
While the touchscreen is the prevailing interaction method, the usage of further modalities broadly varied between participants. One participant used vibration and notification lights as preferably used output; another participant did not use vibration at all, but heavily relied on speech input (although only in private space). A third participant neither used sound nor vibration, but relied on screen notifications, keeping the phone next to him on the desk. This indicates that output modalities should account for diverse preferences and defaults. Further, the focus group revealed that modalities are rarely changed, but that participants maintain “default settings” which they alter only in certain situations. One subject enabled sound only when expecting an important call, while another had the phone in ringing mode as default, but muted the phone in silent environments (e.g., the library). This shows that defaults are interpreted differently.

It also turned out that when speaking of modalities, participants had mostly output in mind. In terms of input methods (other than the touchscreen), participants stated to use device buttons to control the music application or to decline calls, and voice input to set a timer or perform a search query. Hence, input modalities are rather task-specific; therefore a frame-work should support the implementation of novel modalities for individual purposes.

\subsubsection{Identified Problems in Status Quo}
Subjects stated that alternative input methods (i.e., other than standard touch interaction) either do not work reliably enough (speech) or are not implemented consistently (e.g., tap pattern gestures or gaze-based interaction, as offered by some Samsung phones). Participants desire that interaction methods should be more consistent and work in a similar manner in all applications.
One participant stated that he often forgot to revoke a modality change, which could result in unwanted situations (e.g., the phone rings in a silent environment). This fear was confirmed by almost all subjects and pointed out as reason for the presently rare changes of modalities. They would welcome a system that automatically selects modalities, but would like to retain control and be able to override the system behavior. Participants also missed a solution similar to ``profiles'' (as available on earlier Nokia phones), so that a rule-based modality switching approach seemed attractive, especially using context information as a basis for decisions. Participants supposed that most situations could possibly be covered by a small number of rules. Users could then define their own modality settings, which are activated by conditions (such as location, time, etc.) and revoked when the conditions do not apply any more. Participants supposed that rules would be often initially set and later rarely modified. They pointed out that users always need the possibility to override settings when a rule becomes active. In order to prevent forgetting, another idea would be a time limitation for manual settings (e.g., one hour or until the meeting is finished).

\subsection{Developer's Perspective: Expert Interviews}
For the programmer's perspective, we interviewed three software developers involved in mobile application development and asked how satisfied they were with the current tool support for creating multimodal applications. We also were interested in their expectations and wishes for future tools regarding their programming needs.

The following issues were mentioned (summarized and aggregated): Implementing contextual behavior requires the use of different APIs (e.g., Location API, Sensor API, etc.), entailing redundancies and re-implementations of similar pieces of code, as well as heterogeneous ways of accessing data. While for example sensor or location updates are listener-based (push principle), other data must be checked manually (pull principle). To realize context-sensitive behavior based on both push- and pull-based data, interfaces/wrappers need to be created, which adds significant overhead. Developers would appreciate a unified structure for all types of context information, as well as encapsulation of frequently used functions, hiding complexity. In terms of input, novel interaction methods currently have to be designed and implemented from scratch. Especially for rapid prototyping, building blocks would speed up the creation of functional prototypes. In terms of output, it is effortful to include multiple modalities in an application, as each additional output modality must be implemented separately. If developers could in their implementation abstract from the information to be communicated and the method/channel over which it is transported, they were more likely to implement multiple modalities, contributing to more usable, natural, intuitive, and efficient applications.

\subsection{Summary of Requirements}
Based on focus group and expert interviews, we can summarize the following requirements:
\begin{itemize}
\item A holistic system for mobile multimodal interaction should cover both multimodal input (in terms of natural interaction) and output (in terms of context-based modality switches).
\item Defining multimodal behavior based on rules is closest to the human understanding of automated switching and might thus be an adequate underlying model.
\item The system must be flexible enough for the heterogeneous preferences in terms of situational modality settings and defaults.
\item The underlying programming framework should save time, reduce code redundancy, ease the access of contextual information, and abstract from information representations (so that a unit of information can be easily communicated by different modalities).
\end{itemize}

\section{Discrimination from Related Work}
For rapid development of context-sensitive applications in the research context, several toolkits and frameworks have been presented, starting with the \emph{Context Toolkit} by Dey et al.~\cite{dey2001}.
The \emph{Subtle} toolkit by Fogarty and Hudson \cite{fogarty2007} supports the development of applications that use sensor-based statistical models. Thereby, contextual information can be used to adapt certain settings automatically. With our \emph{M3I} framework, we focus on \emph{rule-based} wiring of functionality, as often the impacting factors for a desired action are quite clear. There exist alternative approaches (e.g., support vector machines, Bayesian or neural networks), but machine learning often makes it hard to understand \emph{why} a certain action has been performed. 
Some toolkits address particular use cases, e.g.~physical mobile interaction \cite{pmif2008} or proxemic interaction \cite{marquardt2011}. 
Du and Wang \cite{du2008} developed a model and implementation framework to simulate context events with Symbian phones.
Schuster et al.~used JCOP (Context-Oriented Programming) extensions to the Java language that support context-dependent execution for Android programming \cite{schuster2011}, which, however, complicates a quick integration in any existing Android project and working environment.

Preceding toolkits and frameworks for multimodality that have been presented in research are often focused and confined to specialized use cases, e.g., speech and gesture interaction with large screens \cite{krahnstoever2002real}, or multimodal interaction on a PC \cite{bourguet2002toolkit, flippo2003framework}. The generalizability of such approaches is thus limited. The evaluation of multimodal behavior is conducted by e.g. finite automata \cite{cutugno2012multimodal} or state machines \cite{appert2008swingstate}. The latter approach enhances the Java Swing toolkit to facilitate novel input methods (e.g., multi-handed or pressure-sensitive input). Ravindranath et al. \cite{ravindranath2012code} showed a server-based task distribution and coordination approach, which however entails that the systems using it are not fully autonomous. Manca et al. \cite{manca2013generation} present a multimodal web toolkit, however focusing on browser-based applications. These have limited capabilities compared to native applications, e.g., with relation to hardware and sensor access on mobile devices.

For the choice of optimal modalities (using e.g. the visual, haptic, and auditory channel), often also the user’s context, such as time, location, or the social setting, plays an important role. Context determination is supported by different approaches, such as toolkits \cite{dey2001}, middleware \cite{lee2012mobicon} or context-oriented programming extensions on code level \cite{schuster2011}. They facilitate e.g.~tasking applications like Tasker\footnote{\url{https://play.google.com/store/apps/details?id=net.dinglisch.android.taskerm}, accessed June 4 , 2014} that utilize context to change output modalities (for example, mute ringtones in a silent environment).
While this overview does not claim to be complete, existing toolkits and frameworks seem not to match all of the requirements that emerged from the focus group (see ``Requirements Analysis'' section). To the best of our knowledge, the \emph{M3I} framework we present in this paper is currently the only approach that fulfills all of the following points:
\begin{itemize}
\item Holistic support of modeling, prototyping and implementation of multimodal behavior (often, only individual stages, such as the prototyping stage, are supported)
\item Support of both multimodal input and output (other approaches focus either on the interpretation and fusion of input events, or on output modality switching in form of tasking applications)
\item Support of context-driven multimodality (existing approaches are either pure context toolkits, or focus on multimodal interaction without integrating context, at least not using context for both in- and output)
\item Applicability in standard development environments (code enhancements like JCOP \cite{schuster2011} require adaptations such as a special compiler)
\item Applicability with state-of-the-art mobile operating systems like Android (earlier approaches are dedicated to desktop computing or legacy platforms with limited support for context acquisition)
\item Autonomous operation, without server-based evaluation or control through the cloud as e.g. in Code in the Air \cite{ravindranath2012code}
\item Compliance with the human mental model of multimodal behavior (systems that use automata, state machines, etc. might have worse intelligibility than a rule-based system)
\end{itemize}

\section{M3I: Mobile Multimodal Interaction Framework}
\begin{figure}[!t]
\centering
\includegraphics[width=\columnwidth]{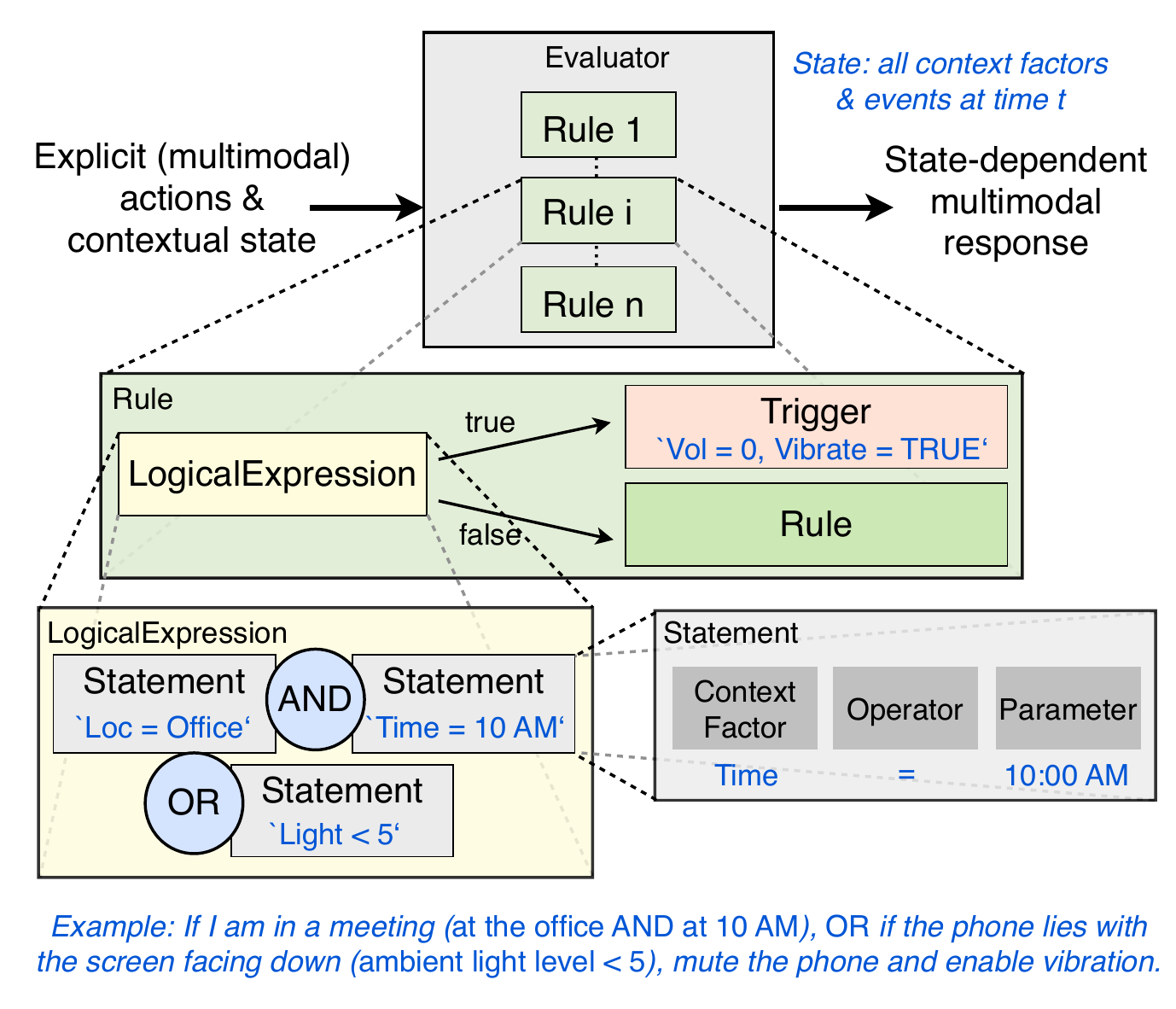}
\caption{General structure of the \emph{M3I} framework: The core is the evaluator, which
evaluates rules that define the system's behavior on explicit user actions and/or implicit context factors and events.
Rules can initiate triggers to directly react on the context and to, e.g., switch modalities, and recursive rules allow implementing complex decision-making.}
\label{fig:state_diagram}
\end{figure}

Our \emph{M3I} framework is implemented as Android library (using the Java language).
Thereby, no special configuration or tools are needed for using it; the library just needs to be referenced from an Android project to use its features.
Figure \ref{fig:state_diagram} shows, on a conceptual level, the structure of the framework and its basic components, which will be detailed in the following.

\subsection{Context Factors and Context Groups}
Context factors are fine-grained pieces of context information that can be used to influence the behavior of a context-sensitive application.
Examples for context factors are the latitude and longitude of the device's position, the charging state of the battery, or the time of day, but also complex information such as the user's current activity.
In the framework, a {\footnotesize\texttt{ContextFactor}} is defined by specifying a {\footnotesize\texttt{ContextGroup}} it belongs to, and the respective context method that provides the {\footnotesize\texttt{ContextFactor}}'s value.
Since Java does not support function references, constants specify the method ID. They are evaluated by an {\footnotesize\texttt{execute()}} method to call the context function appropriately at runtime. 

For example, a context factor of type {\footnotesize\texttt{Boolean}} indicating the device's charging level is defined as follows:

\begin{lstlisting}[language=Java]
FloatContextFactor battLevel =
  new FloatContextFactor(
    new BatteryContext(this),
    BatteryContext.FLOAT_GET_BATTERY_LEVEL);
\end{lstlisting}

Methods for retrieving context information are organized in groups such as {\footnotesize\texttt{OrientationContext}} or {\footnotesize\texttt{LightContext}}, which on their part contain information on the device's orientation or the ambient light level.
Context groups can be seen as collections that organize and provide access to the platform's own methods and routines in a way that eases building up statements and rules (see the following sections).

The framework drastically economizes and simplifies those method calls for the programmer, as it saves all overhead for initializing system services, creating event listeners etc., and makes all functionality accessible by homogenous interfaces.
One particular advantage of the framework is unified handling of synchronous and asynchronous information.
Values available anytime (e.g., the current weekday) can be retrieved with a simple method call,
but dynamic events are normally handled with listeners to receive them asynchronously when they occur (e.g., location changes, sensor events, or touch interactions by the user). 
The framework makes transparent to the developer whether a {\footnotesize\texttt{ContextFactor}} is based on a synchronous or an asynchronous call. It internally creates listeners if required, handles their updates automatically and stores the most recent values. Thus, whenever the contextual state should be determined, a consistent state is available.

\subsection{Abstraction of Information Representation by Triggers}
Triggers define what should happen at a defined contextual state.
They realize modality switches, i.e., they abstract how information shall be represented. For example, a notification can be received as sound or as vibration signal. A row of modalities are directly supported, e.g.~visual (UI changes), haptic (vibration patterns), or auditory (sound playback) responses. Triggers are are not limited to predefined modalities or actions: a {\footnotesize\texttt{MethodTrigger}} calls an arbitrary method to implement custom functionality. With the {\footnotesize\texttt{NullTrigger}} the omission of an action, e.g.~in an \emph{else} branch, can be modeled.
For an example of how triggers are defined, see Demonstration 1 in the \emph{Validation} section.

\subsection{Rule-Based Wiring}
The wiring of input and contextual events and appropriate triggers is realized based on rules.
A {\footnotesize\texttt{Rule}} consists of a {\footnotesize\texttt{LogicalExpression}} which is evaluated to be either true or false, and of actions that define what should happen in either case.
{\footnotesize\texttt{Rule}}s can either call a {\footnotesize\texttt{Trigger}} (as described above), or recursively another {\footnotesize\texttt{Rule}}, which allows the creation of a more complex, nested decision logic.

\subsubsection{Logical expressions}
The simplest form of a logical expression is a {\footnotesize\texttt{Statement}}, which consists of a {\footnotesize\texttt{ContextFactor}} and an {\footnotesize\texttt{Operator}}.
Operators allow value checks 
such as numeric comparisons, within-range-tests, or regular expressions.
Take for example the following statement which checks whether the battery level is above 50\% (it uses the {\footnotesize\texttt{ContextFactor battLevel}} defined in a previous example):

\begin{lstlisting}[language=Java]
Statement isAboveHalfCharged = new Statement(
  battLevel, FloatOperator.greaterThan(50f));
\end{lstlisting}

Using logical operators realized by {\footnotesize\texttt{UnaryExpression}} and {\footnotesize\texttt{BinaryExpression}} classes,
complex terms can be created Although AND, OR, and NOT are sufficient to construct any logical expression according to Boolean algebra laws, XOR, NAND, NOR and XNOR are supported for convenience reasons to implement a decision logic of arbitrary complexity.

The following expression describes the state in which the device is either charged more than 50\% \emph{or} connected to a power plug (in the example, {\footnotesize\texttt{isAboveHalfCharged}} and {\footnotesize\texttt{isPluggedIn}} are previously defined statements):

\begin{lstlisting}[language=Java]
BinaryExpression exp = new BinaryExpression(
  BinaryExpression.EXPRESSION_OR,
  isAboveHalfCharged, isPluggedIn);
\end{lstlisting}

When triggers are defined and the contextual states in which they should be applied have been described by logical expressions, a rule can be created. 

\subsubsection{Rule Evaluation}
The set of active rules is evaluated in the framework's {\footnotesize\texttt{StateMachine}}.
The developer can define the update interval in which the evaluator evaluates the rules and executes the defined triggers.
Thereby, the framework supports time-critical, fast-reacting systems as well as battery-conserving systems where an update e.g.~once per minute is sufficient.
Once a few rules have been defined, the following lines of code activate the framework in any Android application (using a 1000 milliseconds update interval in this example):

\begin{lstlisting}[language=Java]
StateMachine m = new StateMachine(1000);
m.addRule(r_1);
...
m.addRule(r_n);
m.start();
\end{lstlisting}

\subsection{Extensibility}
With a {\footnotesize\texttt{CustomContext}} group, results of any method or callback can be fed into the framework's decision logic when a context factor is not built in directly.
Moreover, by clearly defined interfaces, the framework is easy to extend both on input and output side.
The following simple steps are e.g.~required to add a new context group.
(1) Create a new class implementing the interface {\footnotesize\texttt{IContextGroup}}, and implement all context methods the group should provide, returning a value of arbitrary data type. (2) Define a list of parameters for each function that serves as argument for the {\footnotesize\texttt{execute()}} function of the group.
In a similar manner, new triggers must implement the interface {\footnotesize\texttt{ITrigger}}, which contains a {\footnotesize\texttt{trigger()}} method that performs the desired action.

\subsection{Framework Summary}
The framework currently integrates more than 50 context factors
regarding, e.g., location, ambient noise and light level, device orientation, battery information, proximity information (through NFC, Bluetooth, or Geofence entering/leaving), availability of 3G and WLAN connections, or date and time.
Basic activity recognition and classification routines abstracting from pure sensor readings are already integrated, e.g.~pose classification (in pocket or carried in hand), usage indicators, mode of transportation, vision-based detection (face) etc.

Besides that, explicit interactions, such as physical button presses or touch interactions can be intercepted and combined with implicit contextual information.

On output side, triggers allow controlling a range of modalities and thereby abstracting how information should be represented, based on fine-grained definitions. Examples include visual output (on-screen or via LEDs), sound, brightness setting, or vibration. Further actions include behavioral triggers, e.g.~changing device settings (sync rules, connectivity, screen lock, ...) or custom functions.



%

\section{Validation}
To validate our claim that the framework drastically speeds up the development of context-based multimodal applications, we present three very different applications  demonstrating the wide spectrum of potential use cases for our framework.

\begin{figure}[t]
\centering
\includegraphics[width=0.45\columnwidth]{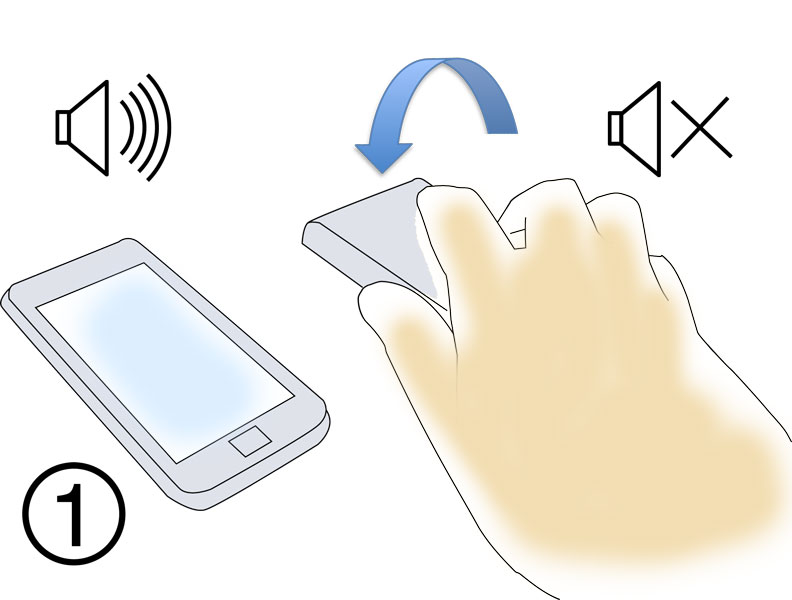}
\hspace{0.5cm}
\includegraphics[width=0.45\columnwidth]{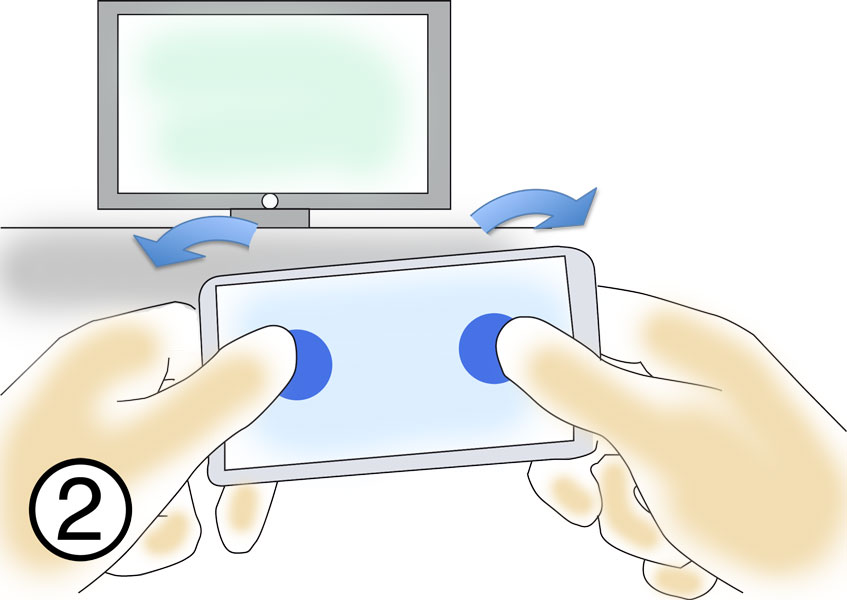}
\caption{Multimodal interaction examples that can be realized with few lines of code using our framework: (1) A turn-to-mute service, switching the phone to silent mode when it is placed upside down (detected by the ambient light sensor). (2) A multimodal game controller using the accelerometer to steer a video game running on a second screen.}
\label{fig:examples}
\end{figure}

\subsection{Demonstration 1: Flip to Mute}
This basic example is a tangible ``mute by flip over'' application for quickly enabling silent mode, e.g.~in a meeting. Instead of manually turning down the volume, the user simply needs to flip over the phone and place it on the table with the display facing down (see Image 1 in Figure~\ref{fig:examples}). A dozen years ago, such context-based telephony applications were prototyped with external sensor modules \cite{schmidt1999}.
In our example, the trigger is realized using the built-in ambient light sensor mounted on the front panel of the phone.
With our framework, this is the code to set up this functionality:

\begin{lstlisting}[language=Java]
// define statement
FloatContextFactor light =
   new FloatContextFactor(lc,
   LightContext.FLOAT_GET_LIGHT_LEVEL);
Statement isUpsideDown = new Statement(light, FloatOperator.smallerThan(5.0f));
// define triggers
AudioTrigger mute = new AudioTrigger(this);
mute.setAction(AudioTrigger.RINGER_VIBRATE);
AudioTrigger ring = new AudioTrigger(this);
ring.setAction(AudioTrigger.RINGER_NORMAL);
// create rule
Rule r = new Rule(isUpsideDown, mute, ring);
\end{lstlisting}

The example does not take into account that the light level can be 0 also at night or when the phone is in a sleeve (it could easily be refined by adding the time of day and the pose as additional context factors), but it demonstrates how few lines of code suffice to realize a multimodal interaction.

\subsection{Demonstration 2: Multimodal Game Controller}
The second example shows how to use a mobile phone as game controller, combining the modalities \emph{touch} and \emph{gestures}. In a racing game running on a second screen, the vehicle can be steered by tilting the phone (see Image 2 in Figure~\ref{fig:examples}).
The angle of the steering wheel is derived from the pose detection context factor, which can directly retrieve the tilt angle of the device.
Touching additional action buttons in the interface on the mobile screen can modify the driving behavior, which demonstrates the framework's capability to seamlessly integrate explicit (button presses) and implicit events (updates from the orientation sensor listener).
The communication to the game running on an Ubuntu desktop PC is realized using ROSjava\footnote{\url{http://code.google.com/p/rosjava/}, accessed June 4, 2014}, a publish/subscribe architecture based on the Robot Operating System\footnote{\url{http://www.ros.org}, accessed June 4, 2014}, a popular architecture for distributed applications in research.

\subsection{Demonstration 3: Mimikry Input Methods}
As multimodal input is (as argued earlier), in contrast to output, task-specific, we implemented three unique methods (\emph{Raise\&Call}, \emph{Press\&Shoot}, and \emph{Pinch$@$Home}) to launch different applications (Phone, Camera, Maps). As our goal was high intuitiveness, we chose metaphors that support the human mental model of performing typical movements with these apps. Each method includes different modalities, like motion gestures (performing a characteristic motion with the device), screen gestures (performing a multi-touch gesture on the screen), or explicit voice or button input (see first three rows of Table 1).
These three input methods were implemented with \emph{M3I} in straightforward manner by a few simple rules (e.g., involving the accelerometer pose as implicit sensor reading and the button press as explicit action for the Press\&Shoot method). Built-in abstractions (predefined device poses like ``upright'', ``lying on the table'', ``display up/down'') additionally accelerated the development of the \emph{Raise\&Call} and \emph{Press\&Shoot} methods. With the rule creation interface shown in Figure 2, end users can easily define similar interaction methods on their own. 

\begin{figure}[!t]
\centering
\includegraphics[width=\columnwidth]{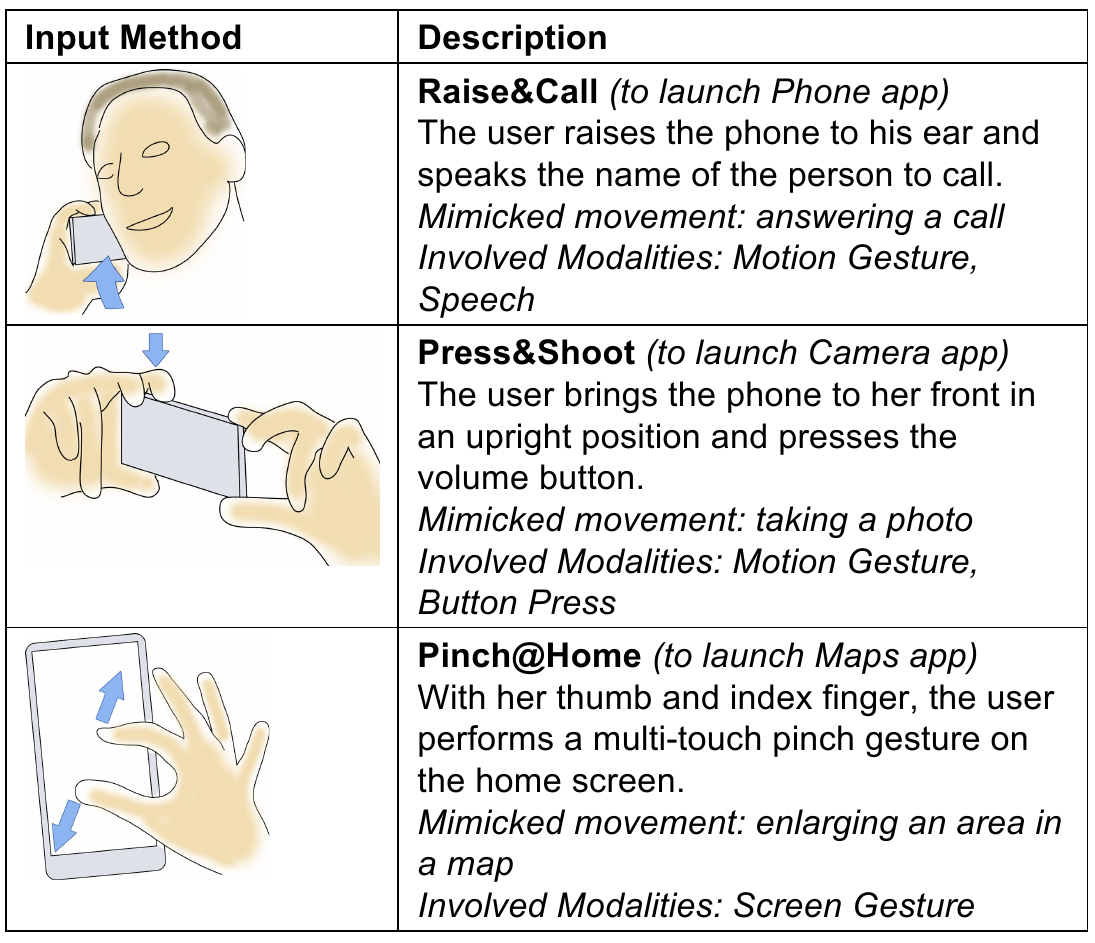}
\caption{Individual intuitive input methods defined with the \emph{M3I} framework to start various applications, mimicking typical gestures and movements when using these apps.}
\label{fig:modality_table}
\end{figure}

\subsection{Demonstration 4: End User Toolkit}
In a final example, we do not focus on a particular scenario, but show an all-purpose application that provides access to the entire functionality of the framework.
As motivated in the introduction, users often have very few and concrete factors that determine a desired modality. However, it is cumbersome to toggle the respective settings each time per hand.
With help of our framework it was little effort to assemble an application allowing end users to automate such actions (similar to, e.g., Llama\footnote{\url{https://play.google.com/store/apps/details?id=com.kebab.Llama}, accessed June 10, 2014}, or Locale\footnote{\url{http://www.twofortyfouram.com}, accessed June 10, 2014} for Android, which are however not focusing on multimodal interaction).
Using a GUI, users can specify their rules conveniently to, e.g., define gestures that trigger certain interactions, contexts in which certain modalities should be enabled, and more.
Based on the GUI, the application creates all required logical expressions, triggers and rules, and runs the evaluator as background service.
Unlike existing applications which confine to a predefined set of simple actions and rules, our framework provides a complex decision logic, high-level contextual abstractions with activity recognition, and it is easily extensible: Recompiling the application with an updated version of the framework directly makes newly added functions accessible.

\begin{figure}[t]
\centering
\includegraphics[width=0.32\columnwidth]{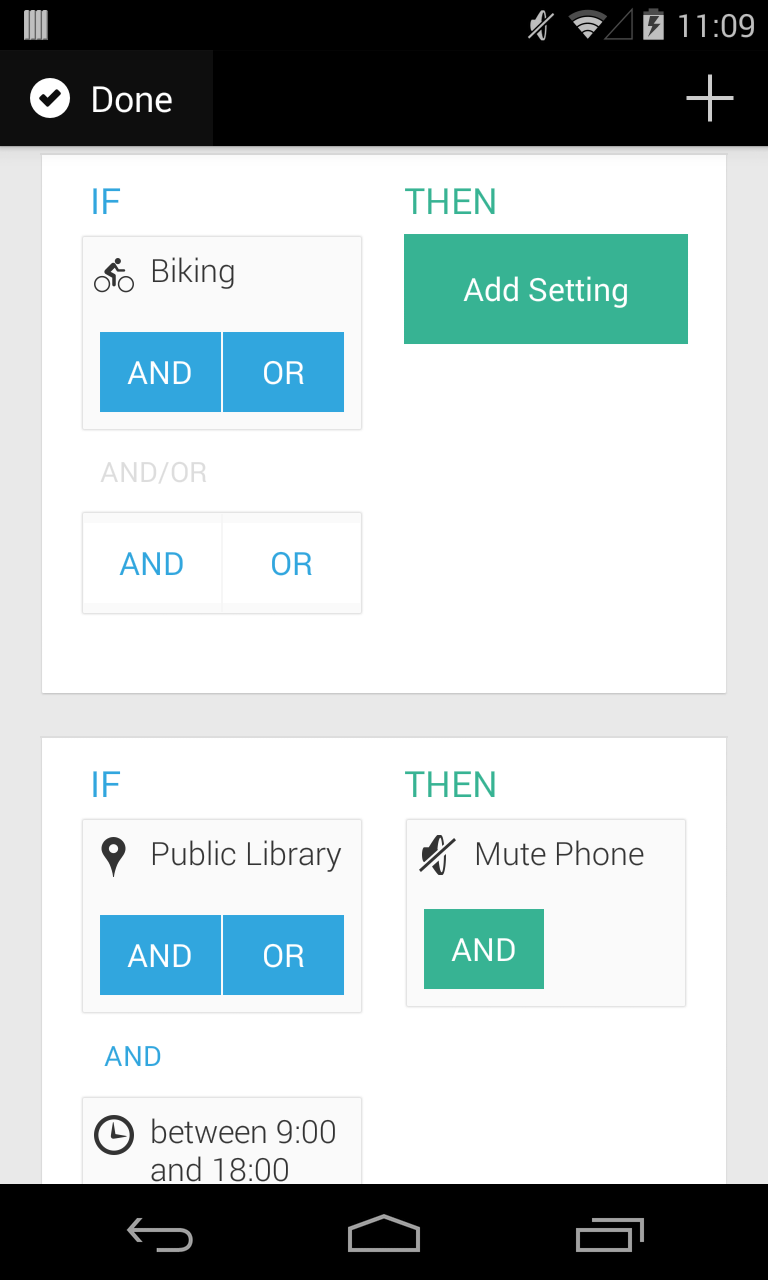}
\includegraphics[width=0.32\columnwidth]{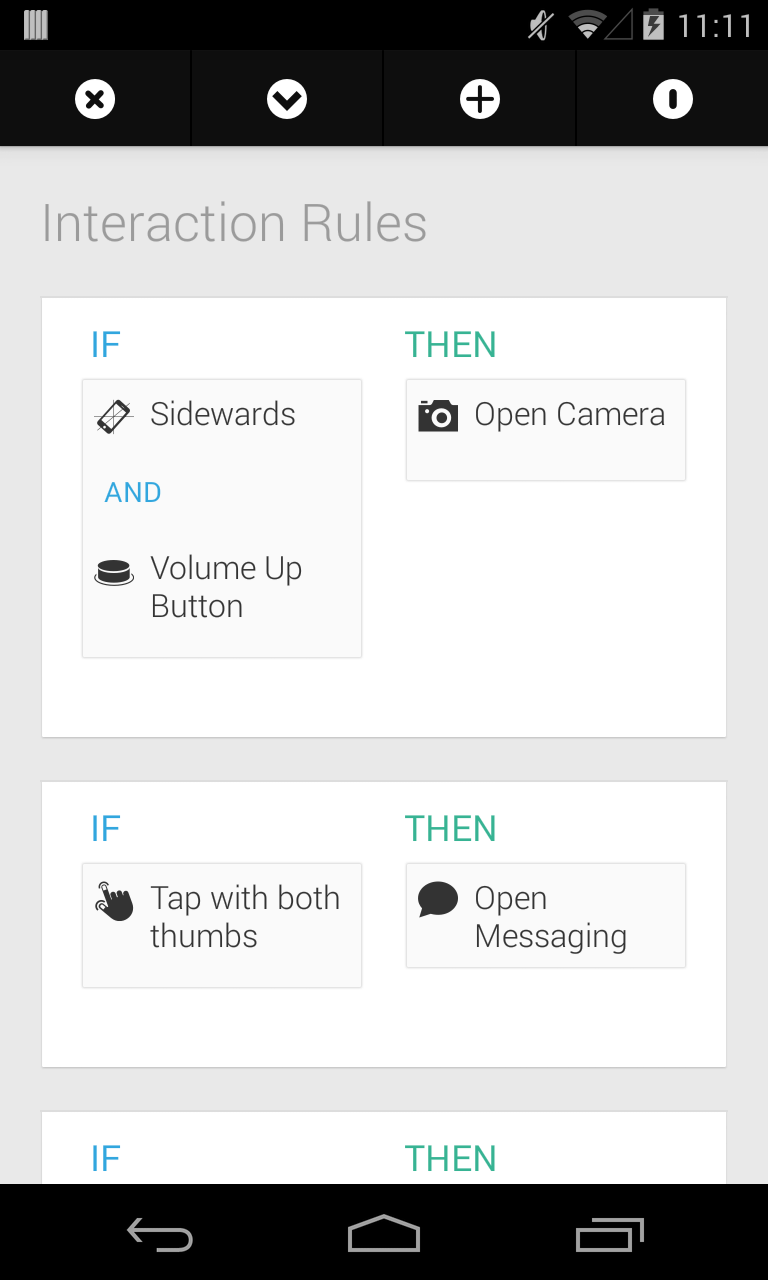}
\includegraphics[width=0.32\columnwidth]{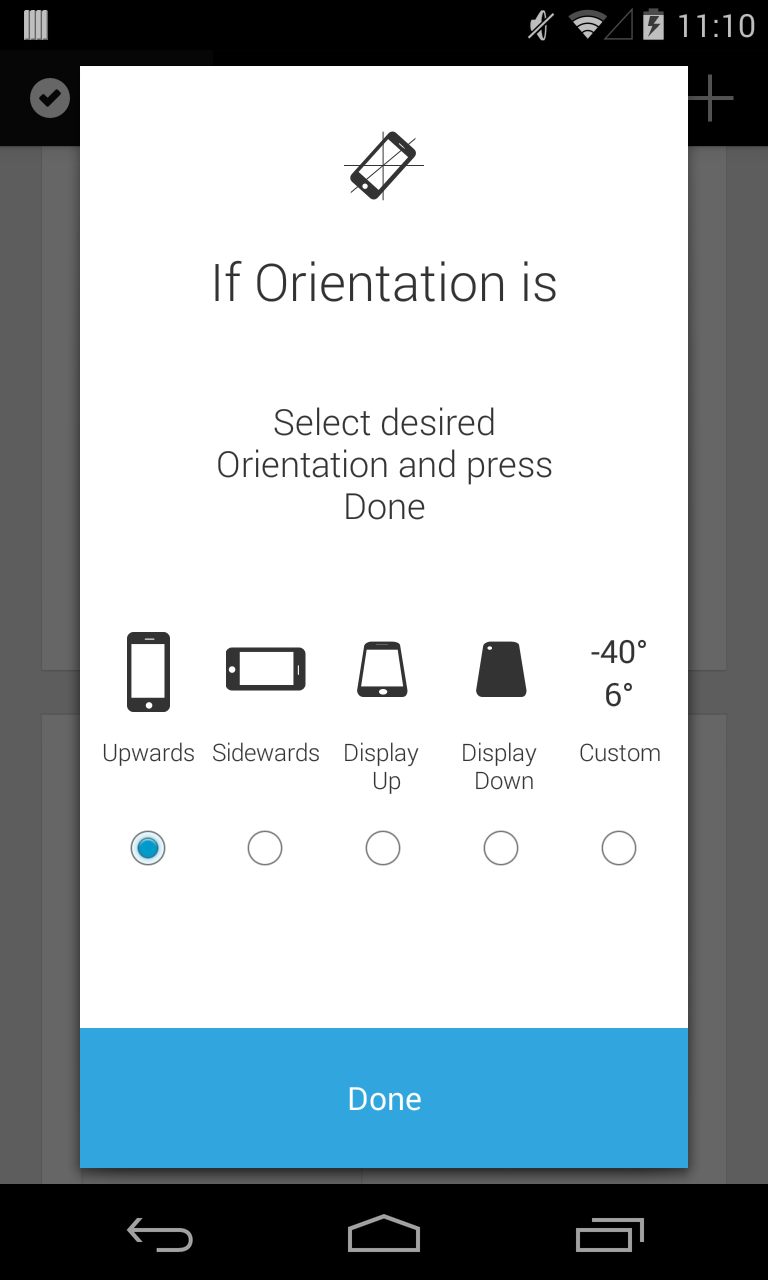}
\caption{With a graphical user interface (GUI), end users can make use of the entire functionality of the \emph{M3I} framework and set up their individual context-defined multimodal behavior. Left: Definitions of new multimodal input methods. Middle: Editing rules for context-dependent output modality switching. Right: Recording orientation (template-based or custom) as part of a multimodal input method.}
\label{fig:enduser}
\end{figure}

Figure \ref{fig:enduser} shows our implementation of a rule-based interface to define system-wide multimodal behavior based on contextual conditions. This interface internally creates {\footnotesize\texttt{LogicalExpression}}s, {\footnotesize\texttt{Trigger}}s and {\footnotesize\texttt{Rule}}s, but is intuitive enough to be operated by end users. In concordance with a suggestion made in the focus group, the system goes back to the previous setting when a rule does not apply any more. This allows using any setting as default. While the left screenshot shows the definitions of some of the interaction methods presented in Demonstration 3, the middle image is an example for the creation of multimodal behavior depending on different contexts, like the mode of transportation or the location. The right screenshot shows the recording of orientation gestures as part of a rule for multimodal interaction. 

\section{Conclusion and Future Work}
The presented \emph{M3I} framework gives researchers and developers a tool at hand that simplifies rapid development and prototyping of multimodal interaction in mobile applications.
While still under active development, we have demonstrated its maturity by the examples in this paper. 

In future work, besides adding more context factors, we will experiment with automated deduction of desired modalities.
The framework could silently observe user behavior, identify patterns and suggest new rules. For example, when a user mutes the device every week at a certain location and time (probably because of a weekly meeting), it could offer to perform this automatically. Unlike classical supervised learning where the learned model is not visible to the user, the rules to be created could be edited and reviewed to establish transparency about an application's multimodal behavior. 

\section*{Acknowledgment}
We thank our student Max Walker for his help with extending the \emph{M3I} framework and with creating the graphical end user toolkit on top of the framework.

%
%
%
%
%
\balance

\bibliographystyle{acm-sigchi}

\begin{thebibliography}{10}

\bibitem{appert2008swingstate}
Appert, C., and Beaudouin-Lafon, M.
\newblock {SwingStates: Adding state machines to Java and the Swing toolkit}.
\newblock {\em Software: Practice and Experience 38}, 11 (2008), 1149--1182.

\bibitem{bourguet2002toolkit}
Bourguet, M.-L.
\newblock A toolkit for creating and testing multimodal interface designs.
\newblock In {\em Proceedings of the ACM User Interface Software and Technology
  Symposium (UIST 2002)}, ACM (2002), 29--30.

\bibitem{cutugno2012multimodal}
Cutugno, F., Leano, V.~A., Rinaldi, R., and Mignini, G.
\newblock Multimodal framework for mobile interaction.
\newblock In {\em Proceedings of the International Working Conference on
  Advanced Visual Interfaces}, ACM (2012), 197--203.

\bibitem{dey2001}
Dey, A.~K., Abowd, G.~D., and Salber, D.
\newblock A conceptual framework and a toolkit for supporting the rapid
  prototyping of context-aware applications.
\newblock {\em Human-Computer Interaction 16}, 2 (Dec. 2001), 97--166.

\bibitem{DBLP:conf/mc/DiewaldMRK12}
Diewald, S., M{\"o}ller, A., Roalter, L., and Kranz, M.
\newblock {DriveAssist - A V2X-Based Driver Assistance System for Android}.
\newblock In {\em Mensch {\&} Computer Workshopband}, H.~Reiterer and
  O.~Deussen, Eds., Oldenbourg Verlag (2012), 373--380.

\bibitem{diewald2012mobilinet}
Diewald, S., M\"{o}ller, A., Roalter, L., and Kranz, M.
\newblock {MobiliNet: A Social Network for Optimized Mobility}.
\newblock In {\em {Adjunct Proceedings of the 4th International Conference on
  Automotive User Interfaces and Interactive Vehicular Applications
  (AutomotiveUI 2012)}} (2012), 145--150.

\bibitem{diewald2013gameful}
Diewald, S., M\"{o}ller, A., Roalter, L., Stockinger, T., and Kranz, M.
\newblock Gameful design in the automotive domain: review, outlook and
  challenges.
\newblock In {\em Proceedings of the 5th International Conference on Automotive
  User Interfaces and Interactive Vehicular Applications (AutomotiveUI 2013)},
  ACM (2013), 262--265.

\bibitem{du2008}
Du, W., and Wang, L.
\newblock Context-aware application programming for mobile devices.
\newblock In {\em Proceedings of the 2008 C3S2E conference}, ACM (2008),
  215--227.

\bibitem{flippo2003framework}
Flippo, F., Krebs, A., and Marsic, I.
\newblock A framework for rapid development of multimodal interfaces.
\newblock In {\em Proceedings of the 5th International Conference on Multimodal
  Interfaces (ICMI 2003)}, ACM (2003), 109--116.

\bibitem{fogarty2007}
Fogarty, J., and Hudson, S.~E.
\newblock Toolkit support for developing and deploying sensor-based statistical
  models of human situations.
\newblock In {\em Proceedings of the SIGCHI Conference on Human Factors in
  Computing Systems (CHI 2007)}, ACM (2007), 135--144.

\bibitem{holleis2006playing}
Holleis, P., Kranz, M., Winter, A., and Schmidt, A.
\newblock Playing with the real world.
\newblock {\em Journal of Virtual Reality and Broadcasting 3}, 1 (2006), 1--12.

\bibitem{jokinen2006user}
Jokinen, K., and Hurtig, T.
\newblock User expectations and real experience on a multimodal interactive
  system.
\newblock In {\em Proceedings of the INTERSPEECH} (2006).

\bibitem{krahnstoever2002real}
Krahnstoever, N., Kettebekov, S., Yeasin, M., and Sharma, R.
\newblock A real-time framework for natural multimodal interaction with large
  screen displays.
\newblock In {\em Proceedings of the 4th IEEE International Conference on
  Multimodal Interfaces (ICMI)}, IEEE (2002), 349.

\bibitem{kranz2005distscroll}
Kranz, M., Holleis, P., and Schmidt, A.
\newblock {DistScroll -- A New One-Handed Interaction Device}.
\newblock {\em 25th IEEE International Conference on Distributed Computing
  Systems Workshops 5\/} (2005), 499--505.

\bibitem{lee2012mobicon}
Lee, Y., Iyengar, S., Min, C., Ju, Y., Kang, S., Park, T., Lee, J., Rhee, Y.,
  and Song, J.
\newblock {MobiCon}: a mobile context-monitoring platform.
\newblock {\em Communications of the ACM 55}, 3 (2012), 54--65.

\bibitem{manca2013generation}
Manca, M., Patern{\`o}, F., Santoro, C., and Spano, L.~D.
\newblock Generation of multi-device adaptive multimodal web applications.
\newblock In {\em Mobile Web Information Systems}. Springer, 2013, 218--232.

\bibitem{marquardt2011}
Marquardt, N., Diaz-Marino, R., Boring, S., and Greenberg, S.
\newblock The proximity toolkit: prototyping proxemic interactions in
  ubiquitous computing ecologies.
\newblock In {\em {Proceedings of the 24th Annual ACM Symposium on User
  Interface Software and Technology (UIST 2011)}}, ACM (2011), 315--326.

\bibitem{moeller2012mobimed}
M\"{o}ller, A., Diewald, S., Roalter, L., and Kranz, M.
\newblock {MobiMed}: comparing object identification techniques on smartphones.
\newblock In {\em Proceedings of the 7th Nordic Conference on Human-Computer
  Interaction: Making Sense Through Design (NordiCHI 2012)}, ACM (2012),
  31--40.

\bibitem{muc2014m3i}
M\"{o}ller, A., Diewald, S., Roalter, L., and Kranz, M.
\newblock {M3I: A Framework for Mobile Multimodal Interaction}.
\newblock In {\em {{Proceedings of Mensch {\&} Computer 2014: Interaktiv
  unterwegs -- Freir\"{a}ume gestalten}}}, Oldenbourg Verlag (M\"{u}nchen,
  2014).

\bibitem{moeller2012mum}
M\"{o}ller, A., Kranz, M., Huitl, R., Diewald, S., and Roalter, L.
\newblock A mobile indoor navigation system interface adapted to vision-based
  localization.
\newblock In {\em Proceedings of the 11th International Conference on Mobile
  and Ubiquitous Multimedia (MUM 2012)}, ACM (2012), 4:1--4:10.

\bibitem{moeller2012trainer}
M\"{o}ller, A., Roalter, L., Diewald, S., Scherr, J., Kranz, M., Hammerla, N.,
  Olivier, P., and Pl\"{o}tz, T.
\newblock Gymskill: A personal trainer for physical exercises.
\newblock In {\em IEEE International Conference on Pervasive Computing and
  Communications (PerCom 2012)} (2012), 213 --220.

\bibitem{MobiDicsVideo2011}
M\"{o}ller, A., Thielsch, A., Dallmeier, B., Roalter, L., Diewald, S.,
  Hendrich, A., Meyer, B.~E., and Kranz, M.
\newblock {MobiDics} -- improving university education with a mobile didactics
  toolbox.
\newblock In {\em {Ninth International Conference on Pervasive Computing
  (Pervasive 2011), Video Proceedings}} (2011).

\bibitem{oviatt99}
Oviatt, S.
\newblock Ten myths of multimodal interaction.
\newblock {\em Commun. ACM 42}, 11 (Nov. 1999), 74--81.

\bibitem{pfleging2012}
Pfleging, B., Kern, D., D\"{o}ring, T., and Schmidt, A.
\newblock Reducing non-primary task distraction in cars through multi-modal
  interaction.
\newblock {\em it - Information Technology 54}, 4 (2012), 179--187.

\bibitem{ravindranath2012code}
Ravindranath, L., Thiagarajan, A., Balakrishnan, H., and Madden, S.
\newblock Code in the air: simplifying sensing and coordination tasks on
  smartphones.
\newblock In {\em Proceedings of the Twelfth Workshop on Mobile Computing
  Systems \& Applications}, ACM (2012), 4.

\bibitem{riedl2013momm}
Riedl, P., Koller, P., Mayrhofer, R., M\"{o}ller, A., Koelle, M., and Kranz, M.
\newblock Visualizations and switching mechanisms for security zones.
\newblock In {\em Proceedings of International Conference on Advances in Mobile
  Computing \& Multimedia (MoMM 2013)}, ACM (2013), 278:278--278:281.

\bibitem{pmif2008}
Rukzio, E., Broll, G., and Wetzstein, S.
\newblock {The Physical Mobile Interaction Framework (PMIF)}.
\newblock Tech. Rep. LMU-MI-2008-2, Ludwig-Maximilians University Munich, 2008.

\bibitem{schmidt1999}
Schmidt, A., Aidoo, K.~A., Takaluoma, A., Tuomela, U., Laerhoven, K.~V., and
  Velde, W. V.~d.
\newblock {Advanced Interaction in Context}.
\newblock In {\em {Proceedings of the 1st Intl. Symposium on Handheld and
  Ubiquitous Computing (HUC 1999)}}, Springer (1999), 89--101.

\bibitem{schuster2011}
Schuster, C., Appeltauer, M., and Hirschfeld, R.
\newblock {Context-oriented Programming for Mobile Devices: JCOP on Android}.
\newblock In {\em Proceedings of the 3rd International Workshop on
  Context-Oriented Programming (COP 2011)}, ACM (2011), 5:1--5:5.

\bibitem{watanuki1994analysis}
Watanuki, K., Sakamoto, K., and Togawa, F.
\newblock Analysis of multimodal interaction data in human communication.
\newblock In {\em Proceedings of the International Conference on Spoken
  Language Processing (ICSLP 1994)}, vol.~94 (1994), 899--902.

\end{thebibliography}

\end{document}